# Measuring the Hole State Anisotropy in $MgB_2$ by Electron Energy-Loss Spectroscopy

Robert F Klie, Haibin Su, Yimei Zhu, and James W Davenport

Brookhaven National Laboratory, Upton, NY 11973

Juan-Carlos Idrobo, Nigel D Browning

University of Illinois at Chicago, Chicago, IL 60607-7059

Peter D Nellist

Nion Co, Kirkland, WA 98033

We have examined polycrystalline $MgB_2$ by electron energy loss spectroscopy (EELS) and density of state calculations. In particular, we have studied two different crystal orientations, [110] and [001] with respect to the incident electron beam direction, and found significant changes in the near-edge fine-structure of the B K-edge. Density functional theory suggests that the pre-peak of the B K-edge core loss is composed of a mixture of $p_{xy}$ and $p_z$ hole states and we will show that these contributions can be distinguished only with an experimental energy resolution better than 0.5 eV. For conventional TEM/STEM instruments with an energy resolution of ~1.0 eV the pre-peak still contains valuable information about the local charge carrier concentration that can be probed by core-loss EELS. By considering the scattering momentum transfer for different crystal orientations, it is possible to analytically separate $p_{xy}$ and $p_z$ components from of the experimental spectra With careful experiments and analysis, EELS can be a unique tool measuring the superconducting properties of $MgB_2$, doped with various elements for improved transport properties on a sub-nanometer scale.

Since its discovery as a superconductor with a transition temperature of $T_c$=39K [1], studies on $MgB_2$ have focused on gaining a deeper understanding of the mechanisms that are responsible for its high transition-temperature and current density.[2,3] Within the theory of hole superconductivity, it became clear that what drives superconductivity in $MgB_2$ is the transport of hole-pairs through the planar boron $\sigma(p_x p_y)$ orbitals. [2,4] Hence, $MgB_2$ is a highly anisotropic BCS superconductor, with the majority of the charge carriers located in the graphite-like B-plane. In a superconductor material, the presence of grain boundaries, oxygen precipitates/segregates, and doping could affect the hole carrier concentration on the order of the coherence length; therefore changing its superconducting properties. Many studies have shown that the current flow between the grains is not reduced by the grain boundaries and even porous or impure samples exhibit a high overall current density.[5] So far, poly-crystalline bulk samples have shown a wide variety of second phases and impurity segregates, such as MgO, $MgB_6$, and $BO_x$ [6,7,8] while maintaining its superconducting characteristics. Significant efforts have be made to increase the critical current density of the bulk materials by deliberately doping the grains with various elements, such as Y[9], Zr[10], C[11], or O[12,6]. Nano-probe spectroscopy is crucial to understand the effects of these dopants and impurities.

In a recent study of hole states in $MgB_2$, using combined X-ray absorption spectroscopy (XAS) and electron energy loss spectroscopy (EELS) we noticed that although the observed total hole states of the p-orbitals agree well with the density functional theory (DFT) calculations, the individual components of $p_{xy}$ and $p_z$ do not. In the TEM based EELS experiments using different instruments and acquisition conditions (collection angles smaller than 1.2 mrad), a higher pre-peak intensity in the B K-edge was observed when the electron beam is parallel to the ab-plane than along the c-axis [13]. Furthermore, DFT simulations of the spectra with a broadened energy resolution suggest an almost inverse behavior of the pre-peak intensity to our experiments. We attributed this discrepancy to the possible matrix element effects in the calculation[13]. In this paper, we report our analysis of the hole state anisotropy in $MgB_2$ and the correlation between the charge carrier concentration and the pre-peak intensity of the B K-edge in $MgB_2$ by combined EELS in both scanning transmission electron microscopes (STEM) and TEMs, multiple scattering (MS) and first-principles DFT calculations. By comparing the fine structure of the B K-edge derived from the MS simulation with the ab-initio results, taking into account the scattering momentum transfer, we conclude that the MS simulation can reproduce the experimental spectra correctly, as long as the experimental energy resolution is not better than 0.8 eV. Nevertheless, to fully understand the recorded near-edge fine structure it is essential to consult DFT calculations. Hence, the combination of high-resolution EELS and DFT calculations provides the ideal tool to study the influence of secondary phases or bulk dopants on the local charge carrier concentration.

The EELS results were obtained using the JEOL 3000F and the JEOL 2010F[14] operated at 300kV and 200kV, respectively; the spectra from both microscopes are in agreement. The microscopes are equipped with a Schottky field-emission source, an ultra high resolution pole piece, an annular dark-field detector and a post column Gatan imaging filter. The microscope and spectrometer were setup for a convergence angle (α) of 13 mrad and a spectrometer collection angle ($\theta_c$) of 38 mrad for STEM with a probe size of 0.2nm and α =0.7mrad and $\theta_c$=1.0 mrad for TEM with a "parallel illumination". The EELS experimental setup allows us to use low-angle scattered electrons [15,16,17,18], which probes the unoccupied density of states near the conduction band minimum. This is analogous to the near edge X-ray absorption spectroscopy (XAS). It is important to note here that although XAS provides a higher energy resolution (<0.1 eV), the lack of spatial resolution does not allow studying the angular dependence of the near-edge fine structure in the poly-crystalline $MgB_2$ sample. Complementary to parallel illumination, a high localization of the core-loss signal can be achieved with a collection angle larger than 20 mrad to assure high spatial resolution of the EELS spectra.[19]

In the experiment performed here, EEL spectra of the B K-edge are acquired directly from grains in different orientations with an acquisition time of 2s per spectrum. The experimental spectra shown here are a sum of 15 individual spectra, added to increase the signal to noise ratio of the near-edge fine-structure. Further, the spectra are background subtracted and deconvolved with the zero-loss peak to remove the effects of plural scattering from the core-loss spectra, unless otherwise stated.

The MS simulations for EELS spectra are performed using the FEFF8 codes.[20] These simulations are based on a self-consistent real-space multiple-scattering (RSMS) approach that calculates the x-ray absorption near-edge structure (XANES) of the desired core-loss edge using *ab-initio* codes. FEFF8 can be used in arbitrary aperiodic or periodic systems for typically less than 100 atoms, while including full multiple scattering from atoms within a well defined cluster. Higher-order multiple scattering contributions from atoms outside this cluster are also taken into consideration. The near-edge fine-structure of the simulated spectra converge for cluster-sizes of 74 atoms, and all the subsequent MS-simulations are performed on clusters of this size.

The DFT simulations for density of the states were performed using the full potential linear augmented plane wave (FLAPW) method as embodied in the Wien2k code.[21] The generalized gradient approximation of Perdew-Burke-Ernerhof was used for the exchange-correlation potential.[22] Within the muffin-tin spheres ($r_{MT,Mg}$=2 a.u., $r_{MT,B}$=1.6 a.u.), lattice harmonics up to l=6 were included for both the density and potential. The cutoff of the plane wave expansion was given by $R_{MT}*K_{max}\approx 8$. Self-consistency was achieved by demanding the convergence of the total energy to be smaller than $10^{-5}$ Ry/cell. For these calculations the self consistent ground state potential was used. Calculations using a potential with a core hole (corresponding to the final state) reveal a shift of the features to lower energy but not a qualitative change in the spectrum.

It is very important to note here that all the EEL spectra, acquired under the condition reported above contain electronic transitions from a range of incident and collecting angles. Therefore, the momentum-transfer contributions in a uniaxial material, both parallel and perpendicular to the incident electron beam must be taken into consideration to properly understand the spectral features.[23] Since the exact contributions from each component is determined by the crystal symmetry, the specimen orientation, the probe convergence angle and the collection angle, the ratio of the individual components can be calculated by the energy-loss function equation for crystals with a hexagonal symmetry [23]:

$$E = (\varepsilon^\perp - \varepsilon^\parallel)\left(\frac{\theta_c^2}{\theta_c^2 + \theta_E^2}\right) + (\varepsilon^\perp + \varepsilon^\parallel)\ln\left(1 + \frac{\theta_c^2}{\theta_E^2}\right) \text{ for } v \perp c \quad (1a)$$

$$E = (\varepsilon^\parallel - \varepsilon^\perp)\left(\frac{\theta_c^2}{\theta_c^2 + \theta_E^2}\right) + \varepsilon^\perp \ln\left(1 + \frac{\theta_c^2}{\theta_E^2}\right) \text{ for } v \parallel c \quad (1b)$$

where $\theta_c$ is the scattering angle, $\theta_E$ is the characteristic scattering angle for a given energy loss ($\theta_E = \Delta E/2E$, where $E$ is the beam energy and $\Delta E$ is the energy loss), $\varepsilon^\perp$ and $\varepsilon^\parallel$ are the perpendicular and parallel components of the dielectric function, $v$ is the direction of the in coming electron beam. $\theta_E$ is energy dependent but its variation is less than 1% for $\Delta E$ in the range of 187 to 225 eV, and it can therefore be considered constant at $\theta_E = 0.47$ *mrad* for 200 keV ($\theta_E = 0.3$ *mrad for* 300 keV).

Figure 1 shows a graphic representation of equation 1a) and 1b). Here, only the parallel components of the dielectric function as a fraction of the total spectral weight are plotted. Three distinct regimes can be identified in this plot; at small collection angle (i.e. smaller than $\theta_C < 2\theta_E$) the forward scattered components are primarily recorded, at the "magic angle" conditions (i.e. $\theta_C \sim 3\theta_E$) the fraction of the parallel component is independent of the grain orientation, and at large collection angle (i.e. $\theta_C > 5\theta_E$) only minor changes occur when increasing the collection angle. It should be noted here that due to the very small characteristic scattering angle of boron, most STEM/TEM EELS experiments are performed with a collection angle much larger than the $\theta_E$ to couple a reasonable beam current into the spectrometer. A direct interpretation of the fine structure, especially the B-Kedge pre-peak in terms of hole-state anisotropy and charge carrier concentration is therefore not possible and comparison to electronic structure calculations is necessary.

Table 1 shows the different scattering contribution for various collection angles for 200 keV incident electrons. We note that only for collection angles smaller than $\theta_E$, the momentum transfer parallel to the incident beam is dominant. For $\theta_C =38\ mrad$ (STEM setting) the parallel component $\varepsilon^{\|}$ as a fraction of the total spectral weight is 44% in the [110] direction and 11% in the [001] direction, while for $\theta_C =1.8\ mrad$ (TEM setting), $\varepsilon^{\|}$ only contributes 1/3 of the total spectral weight, regardless of the crystal orientation.

The B K-edge core-loss EEL spectrum from a grain in the [110] orientation is displayed in Figure 2a; Figure 2b contains the EELS from a grain in the [001] orientation. The B K-edge onset for the two orientations was determined to be at $(187.5 \pm 0.7)$eV , which is good in agreement with previous work by different techniques including x-ray photoemission spectroscopy (XPS), XAS and DOS calculations.[24] Two main features can be observed directly from the experimental spectra. Firstly, the pre-peak of the B K-edge near the threshold energy is significantly higher in the [110] orientation spectra than in the perpendicular direction. It is believed that this peak is formed by the preferential scattering into the empty ($p_{xy}$) states, which would indicate a higher hole concentration in the ab-planes. Further, the peak intensity at 192 eV in the [110] spectra is larger than in the [001] orientation. This peak was reported to represent transitions to the localized empty $\pi^*(p_z)$ states. [13, 24]

The corresponding MS-simulated spectra for the given sample orientation are also shown in Figure 2. The edge onset for the simulated spectra is shifted to match the experimental core-loss edge onset, and the intensity of the simulated spectra is normalized to the continuum 60 eV beyond the edge onset. The intrinsic energy resolution of the MS-simulation is ~0.8 eV and the spectra displayed here are not broadened to match the experimental resolution. Each spectrum contains the parallel and the perpendicular scattering contributions with respect to the incident beam direction and the sample orientation, according to Equation 1. More specifically, the MS-spectrum in Figure 2a contains 56% of the scattering polarized in the [110] direction and 44% polarized in the [001] direction and the [110] direction contributes 89% to the total weight of the spectrum and 11% stem from the [001] direction in Figure 2b. A comparison between the experimental and the simulated spectra for both specimen orientations shows a remarkable match of the individual peak positions The simulated spectra also show the decrease in the pre-peak for the [001] orientation, although the peak intensity at 192 eV is overestimated for both directions. The fact that the intensities between 198 -225 eV are underestimated in the simulated spectra could possibly be explained by multiple scattering effects in the experimental spectra, and by resonant transitions that cannot be properly simulated.

Figure 3 shows the reconstructed polarized contribution in the [110] and the [001] direction, respectively. The experimental spectra can be separated into the polarized contribution according to Equation 1. These spectra then represent the ones acquired with a very small collection angle, so that only the components parallel to the incoming beams are selected. The comparison with the simulated MS-spectra shows that for the [110] spectrum the peak positions are very nicely reproduced but the intensity of the peak at 192 eV is overestimated. Nevertheless, it is remarkable that even the small pre-peak at 186 eV appears in both the experimental as well as in the simulated spectrum. This pre-peak is directly related to the $p_{xy}$-hole concentration in the ab-plane of $MgB_2$. Although the intensity of this pre-peak appears smaller in the reconstructed spectrum, which can be attributed to the inferior energy-resolution, for the first time the $p_{xy}$ and $p_z$ contributions to the pre-peak intensity can be analytically separated. Figure 3b shows the reconstructed spectra in the [001] direction experiment and theoretical calculations. Here, the first peak at 187 eV corresponds to the $p_z$-hole peak intensity, which contributes significantly to the pre-peak in both, the [110] and the [001] experimental spectra. The second shoulder at 204 eV does not appear in the MS-simulation and also the drop in intensity at 216 eV is not seen in the simulated spectra. The differences in the individual spectra will be the subject to further studies.

Figure 4a) shows the result of the DFT simulation for directions parallel to the ab-plane broadened by 0.5 eV and the corresponding reconstructed spectrum. The pre-peak at 186 eV, which represents transitions into the $p_{xy}$-hole-states can clearly be identified in both spectra. The peak-intensities at 192 eV and 203 eV are more accurately reproduced in the DFT simulations than in the MS-calculations and also the decrease in edge intensity at 206 eV matches the experiment. Although the DFT simulation produces a much higher number of distinct peaks that are not distinguishable in the experimental spectrum, the general shape of the (110)-polarized spectrum is reproduced.

The simulation for grains in the [001]-orientation (Figure 4b) shows a high intensity at the edge onset, with a sharp peak at around 188eV. This peak cannot be seen in the experimental spectrum, but the shoulder at the edge onset might indicate it presence. The general shape of the calculated spectrum between 194 eV - 202 eV fit remarkably with the experiment, and even the peaks at 204 eV and 209 eV appear both in the calculations and in the experiment. Overall, the simulations for both orientations show agreement with the experiments, and allow a more precise interpretation of the core-loss near edge fine-structure.

Figure 5 shows an experimental spectrum acquired with the Nion VG HB501, a dedicated STEM (DSTEM), having a cold-field emission source operated at 100 keV with a nominal energy resolution of 0.4 eV. In addition, this microscope is equipped with an objective lens spherical aberration ($C_s$) corrector[25] that enabled us to achieve a convergence angle of 22 mrad and an electron-probe size of 1.4 Å. The spectrometer collection half-angle was about 30 mrad. The spectrum shown in figure 5 is the sum of four individual spectra, acquired with 2s exposure time and a dispersion of 0.1 eV/pixel. The main difference between this spectrum and the ones shown in figure 1 is that an additional shoulder in the pre-peak can be clearly identified at 189 eV. The DFT results for an arbitrary grain orientation and the two polarized components, all energy broadened by 0.5 eV are also shown in figure 5. The comparison with the DFT simulation clearly shows that a splitting of the pre-peak is expected, due to the individual contributions from the $p_{xy}$ and the $p_z$ hole-states. The intensity at the edge-onset (186 eV) can be now identified as the sum of both the $p_{xy}$ and the $p_z$ states, whereas the second shoulder (189 eV) is caused only by transitions into the $p_z$ states. It has to be stated here that the experimental spectra are not corrected for plural scattering contribution, hence only the fine-structure close to the edge-onset can be compared directly with the calculations. Nevertheless, a match between the experimental results and the calculation was observed, suggesting the increased energy resolution of the experiments will provide valuable insights into the formation mechanisms of the B K-edge near-edge fine-structure.

In this paper, the near-edge fine structure of the B core-loss in $MgB_2$ was successfully simulated by both MS and ab-initio calculations for two different crystal orientations. The changes in the pre-peak intensity that were believed to be directly correlated to the hole-concentration in the ab-plane of $MgB_2$ were reproduced in both cases. It was shown that the pre-peak intensity is composed of two contributions, namely the $p_{xy}$-hole peak at 186 eV and the high-intensity $p_z$-hole peak at 189 eV. In the experimental spectra (Figure 2) it is impossible to distinguish these two contributions directly with an energy resolution of ~1 eV. However, by decomposing the experimental spectra into its two basis contribution, the $p_{xy}$- and the $p_z$-hole-peak can be separated. The MS- and DFT calculations may be used to confirm and predict the general features seen in the EELS experiment.

In preparing submission of this manuscript, we noticed a newly published orientation dependent EELS study of $MgB_2$ [26]. Although similar methods were used to measure and calculate the B K-edge and Mg L-edge fine-structure, the authors were not able to conclusively link the changes in the B K-edge pre-peak fine-structure to the charge carrier anisotropy. Nevertheless, the experimental results presented in that report are in agreement with the simulation described here, considering the smaller collection angle used. The paper[26] provides further proof that the momentum transfer theory[23] described here, can be successfully applied to a wide range of convergence and collection angles.

In conclusion, high-resolution EELS in electron microscopes, as reported in this paper, allows measuring the anisotropy of the charge carrier concentration in $MgB_2$, when the proper experimental acquisition conditions are taken into consideration. For most experimental acquisition conditions a mixture between the parallel and perpendicular scattering contributions is expected. Furthermore, using a STEM/TEM with an energy resolution of poorer than 1 eV, the unoccupied $p_{xy}$-states close to the Fermi-level cannot be separated from the $p_z$-states in B K-edge pre-peak. Our simulation, as well as experiment (Fig.5), suggest that the $p_z$-states are peaked at about 2.3eV higher than the $p_{xy}$-states in the B K-edge. These two states can only be separated by EELS with an improved energy-resolution. Direct measurements of the $p_{xy}$-charge carrier concentration with a sub-nanometer accuracy will then be possible. In the future, the effects of zone-axes probing, electron beam irradiation and violation of the dipole selection rule for large scattering angles on the fine structure have to be evaluated. Further theoretical studies are also required to understand the cause of the discrepancy in the fine details of core-loss spectra between the calculations and experiments. Then, the effects of anion and cation doping on the charge carrier can be analyzed using the techniques described in this paper and possible pinning centers in this newly discovered superconductor can be identified directly.

The authors would like to thank Dr. G. Schneider for his valuable contributions to the ab-initio simulations. This work is supported by the U.S. Department of Energy, Division of Materials Sciences, Office of Basic Energy Science, under Contract No. DE-AC02-98CH10886 and DOE FG02 96ER45610 . The experimental results were obtained in parts on the JEOL 2010F operated by the Research Resources Center at UIC and funded by NSF, and on the VG HB501 located at Nion Co, Kirkland, WA.

**Figure 1:** Contribution of the parallel component in the total EEL spectrum for the grain orientations, where the c-axis is parallel (a) and perpendicular (b) to the incoming electron beam direction *v*.

**Figure 2:** a) Experimenal EEL spectrum and MS-simulation from a grain for the electron beam direction parallel to the crystal [110] orientation; b) the [001] orientation. (E-$E_f$) denotes the energy above the edge onset

**Figure 3:** a) Reconstructed EEL spectrum and MS-simulation polarized in the [110] orientation ($\varepsilon^{\perp}$); b) in the [001] orientation ($\varepsilon^{\parallel}$)

**Figure 4:** *DFT* simulation for the a) [110] b) [001] polarization including the results of an experimental broadening of 0.5 eV

**Figure 5:** Experimental spectrum with a nominal energy resolution of 0.4 eV. In comparison the corresponding *DFT* simulation and the individual components are also shown.

**Table 1:** Momentum transfer for the parallel ($\varepsilon^{\parallel}$ parallel to the c-axis) and perpendicular ($\varepsilon^{\perp}$ perpendicular to the c-axis) components of the dielectric function with respect to the incident-beam direction *v* for different collection angle and for two crystal orientations.

**Figures:**

Figure 1:

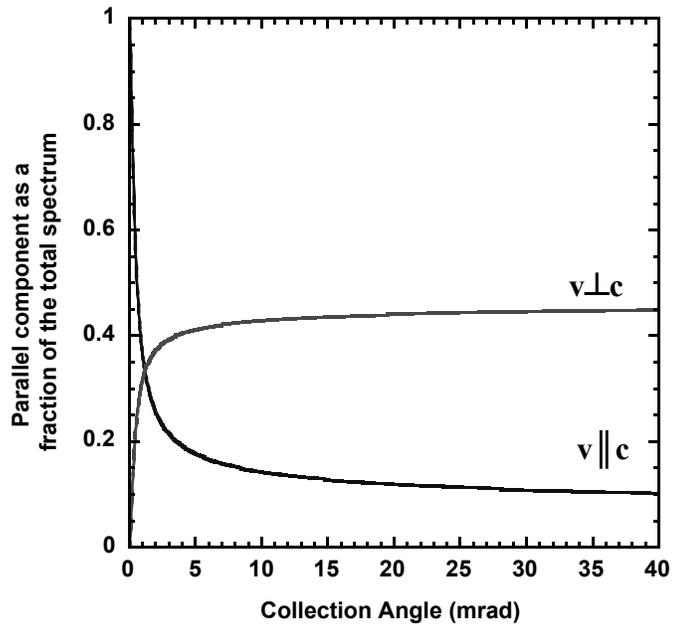

Figure 2a

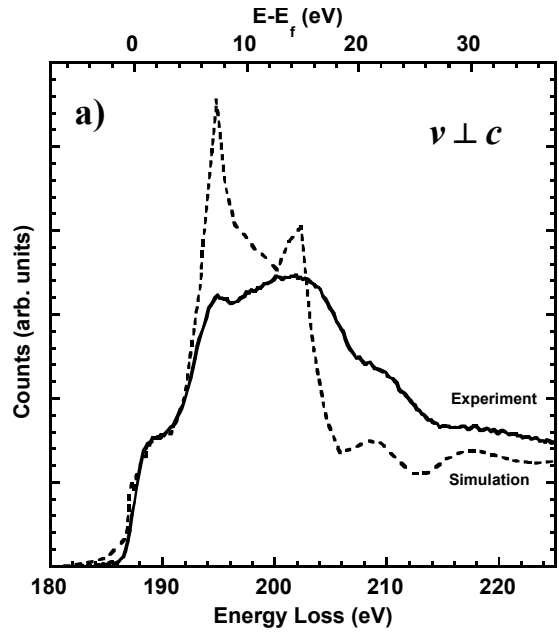

Figure 2b

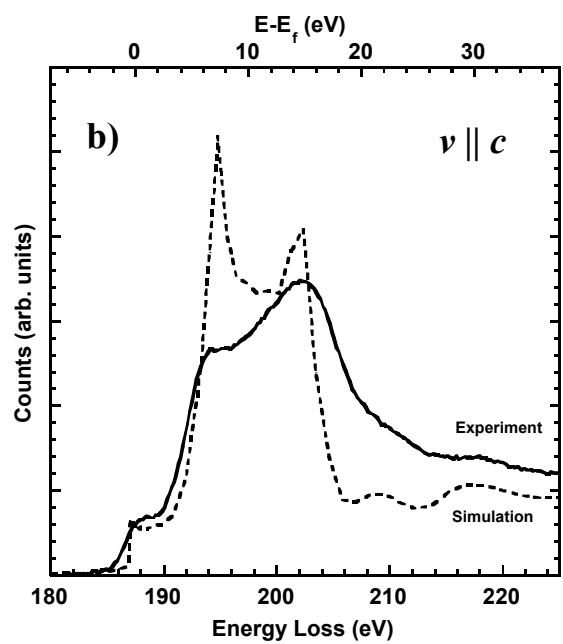

Figure 3a:

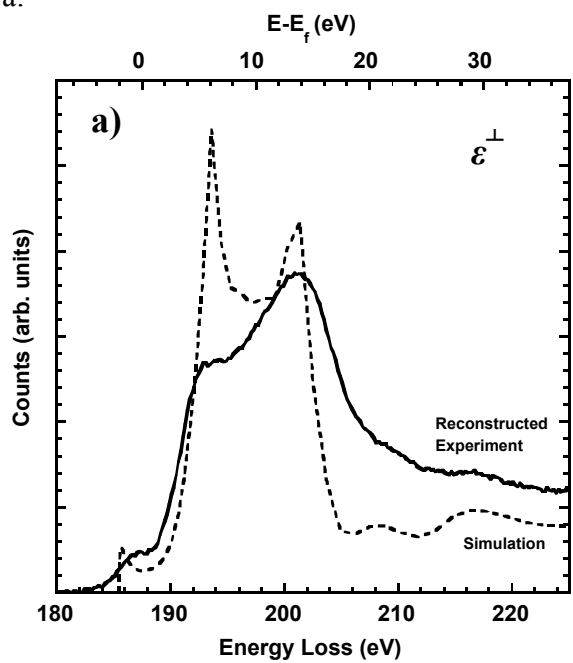

Figure 3b:

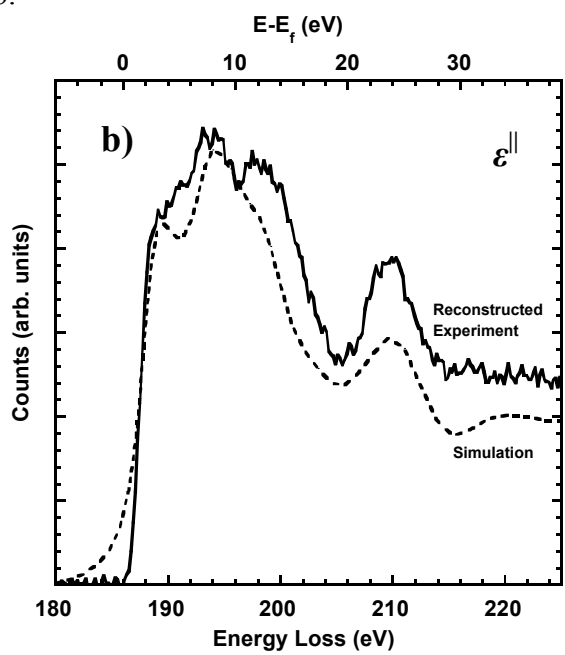

Figure 4a:

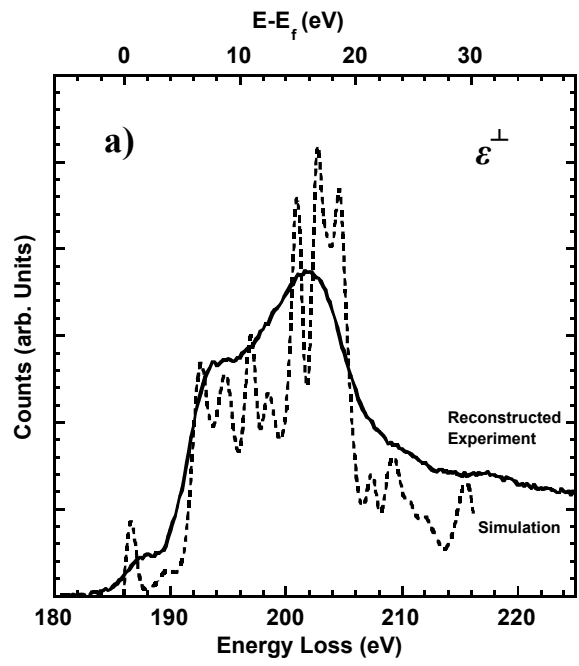

Figure 4b:

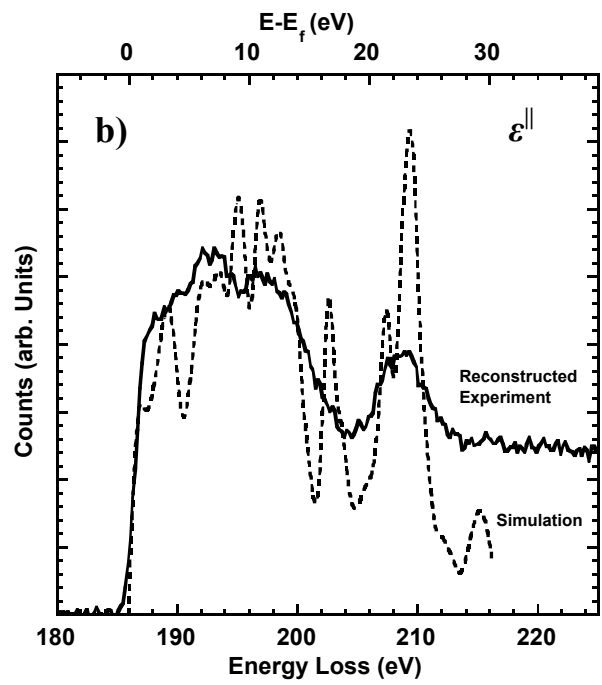

Figure 5:

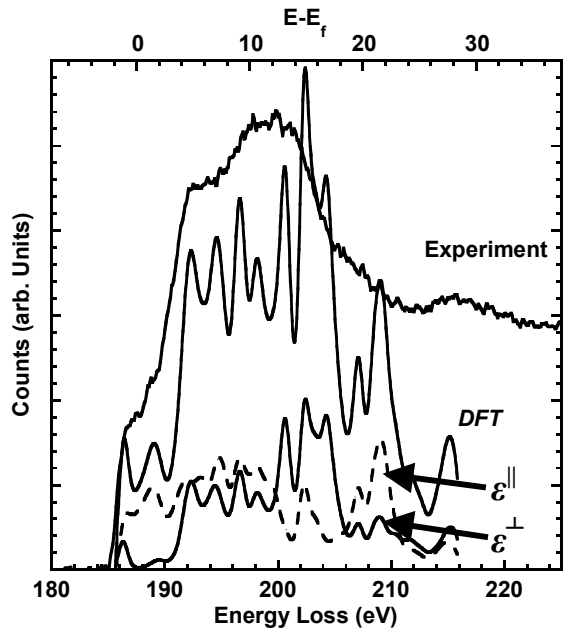

Table 1:

| $\theta_C$/mrad | v∥c | | v⊥c | |
|---|---|---|---|---|
| | $\varepsilon^{\parallel}$ | $\varepsilon^{\perp}$ | $\varepsilon^{\parallel}$ | $\varepsilon^{\perp}$ |
| 0.1 | 0.95 | 0.05 | 0.03 | 0.97 |
| 0.47 | 0.72 | 0.28 | 0.14 | 0.86 |
| 1.8 | 0.33 | 0.67 | 0.33 | 0.67 |
| 5 | 0.21 | 0.79 | 0.39 | 0.61 |
| 38 | 0.11 | 0.89 | 0.44 | 0.56 |